\documentclass[conference]{IEEEtran}
\IEEEoverridecommandlockouts
\usepackage{cite}
\usepackage{amsmath,amssymb,amsfonts, mathrsfs}
\usepackage{algorithmic}
\usepackage{graphicx}
\usepackage{textcomp}
\usepackage{xcolor}
\usepackage{multirow}
\usepackage{url}
\usepackage{bbm}
\usepackage{booktabs}
\usepackage{diagbox}
\usepackage{flushend}

\def\BibTeX{{\rm B\kern-.05em{\sc i\kern-.025em b}\kern-.08em
    T\kern-.1667em\lower.7ex\hbox{E}\kern-.125emX}}

\newcommand{\minew}[1]{{\color{red}{#1}}}

\begin{document}

\title{Transferable Watermarking to Self-supervised Pre-trained Graph Encoders by Trigger Embeddings}

\author{
\IEEEauthorblockN{Xiangyu Zhao}
\IEEEauthorblockA{\textit{School of Communication \& Information} \\
\textit{Engineering, Shanghai University}\\
Shanghai 200444, China \\
zzxxyyy123@shu.edu.cn}
\and
\IEEEauthorblockN{Hanzhou Wu$^*$\thanks{$^*$Author to whom any correspondence should be addressed.}}
\IEEEauthorblockA{\textit{School of Communication \& Information} \\
\textit{Engineering, Shanghai University}\\
Shanghai 200444, China \\
h.wu.phd@ieee.org}
\and
\IEEEauthorblockN{Xinpeng Zhang}
\IEEEauthorblockA{\textit{School of Communication \& Information} \\
\textit{Engineering, Shanghai University}\\
Shanghai 200444, China \\
xzhang@shu.edu.cn}}

\maketitle

\begin{abstract}
Recent years have witnessed the prosperous development of Graph Self-supervised Learning (GSSL), which enables to pre-train transferable foundation graph encoders. However, the easy-to-plug-in nature of such encoders makes them vulnerable to copyright infringement. To address this issue, we develop a novel watermarking framework to protect graph encoders in GSSL settings. The key idea is to force the encoder to map a set of specially crafted trigger instances into a unique compact cluster in the outputted embedding space during model pre-training. Consequently, when the encoder is stolen and concatenated with any downstream classifiers, the resulting model inherits the `backdoor' of the encoder and predicts the trigger instances to be in a single category with high probability regardless of the ground truth. Experimental results have shown that, the embedded watermark can be transferred to various downstream tasks in black-box settings, including node classification, link prediction and community detection, which forms a reliable watermark verification system for GSSL in reality. This approach also shows satisfactory performance in terms of model fidelity, reliability and robustness.
\end{abstract}

\begin{IEEEkeywords}
Watermarking, graph self-supervised learning, graph neural networks, ownership protection, security.
\end{IEEEkeywords}

\section{Introduction}
As an important data structure, graph represents the relationship among elements. Graph data is ubiquitous in the real world, such as social networks and citation networks. With the development of artificial intelligence techniques over the past decade, deep learning models for graphs, i.e., Graph Neural Networks (GNNs), have become the default choice for processing graph data \cite{zhang2020deep, battaglia2018relational, kipf2016semi, xu2018powerful}. Traditionally, GNNs generally rely on (semi-)supervised learning to learn representations of graphs in the Euclidean space with the help of supervision signal from expensive human-annotated data, which often fall short in terms of generalization ability and adversarial robustness \cite{liu2022graph}. To overcome such problems, like in other domains, self-supervised learning (SSL) for graph models, i.e. , Graph Self-Supervised Learning (GSSL), has been developed and can be adopted in various graph-related tasks, such as node classification, edge prediction and graph classification. The performance of GSSL is even comparable with traditional (semi-)supervised learning \cite{caron2021emerging, wu2023brief, liu2022graph}. 

Technically, the main idea of GSSL is to pre-train encoders on handcrafted pretext tasks using only unlabeled data, which helps encoders to learn informative representations of the input data itself. As a result, a well-pretrained encoder can then be combined with various task-specific downstream models (i.e., classifiers) for downstream tasks in applications, where only a small amount of labeled data is needed for adapting the model to the target task domain via fine-tuning. Such a framework relieves the requirement for a large amount of prohibitive human-annotated data and improves the generalization of the resulting combined model. 

However, the strong generalization ability and easy-to-plug-in nature of graph encoders also provide convenience for attackers. For example, an attacker can steal a pre-trained encoder in an open-sourced platform to build his/her own application without notifying the model owner or acknowledging the copyright information, which severely damages the intellectual property (IP) of the model owner. Under such circumstances, the stolen encoder is actually integrated inside the whole model, which brings obstacles for the model owner to claim the rightful copyright. 

To protect deep neural networks from copyright infringement, people typically resort to DNN watermarking techniques \cite{uchida2017embedding, li2021survey}. However, existing watermarking methods for GNNs are designed under the framework of (semi-)supervised learning, which cannot be applied to graph encoders trained in a self-supervised manner \cite{xu2023watermarking, zhao2021watermarking}. To this end, we design a novel watermarking method for node-level graph encoders that can verify the copyright information in black-box settings. Specifically, we inject a `backdoor' into the graph encoder by forcing its output representation of a set of specially crafted trigger nodes as a compact cluster in the embedding space. As a result, when the model is stolen and concatenated with the downstream model for the downstream task, the injected backdoor is inherited by the resulting combined model, which gives the trigger nodes similar predictions with high probability. Such behavior only occurs in watermarked models, which actually forms a 0-bit watermarking in the black-box settings. Experiments show that the embedded watermark can be reliably extracted in black-box settings, which demonstrates the effectiveness of the proposed watermarking framework on protecting graph encoders. 
To summarize, the main contributions of this paper include:

\begin{itemize}
	\item We propose a novel watermarking scheme for node-level GNNs under the self-supervised learning framework. Our work primarily sheds light on the copyright protection of modern GSSL models.
	\item The proposed method embeds the watermark into the embedding space of graph encoders that can be inherited to any models that integrate the protected encoder, which can be verified in black-box settings.
	\item Extensive experiments on various benchmarking datasets with modern GSSL schemes have demonstrated the effectiveness of the proposed watermarking scheme.
\end{itemize}

\section{Background}
\subsection{Graph Neural Networks}
To efficiently and effectively learn the pattern encoded in the topological structure and node attributes in graph data, existing GNN architectures generally follow the message passing framework \cite{gilmer2017neural}, where the representation of each node is derived from its locally neighboring nodes. The representations are further processed by downstream models according to the learning task. Common learning tasks in graph domain include node-level tasks, edge-level tasks and graph-level tasks.

Conventionally, GNNs are trained in an end-to-end manner via (semi-)supervised learning. Taking the typical node classification as an example, given a graph with a set of labeled nodes, the model is trained using labeled nodes so as to accurately predict the labels of the remaining unlabeled nodes \cite{kipf2016semi}. However, this mechanism requires expensive human-annotated data, which gives rise to the self-supervised learning for GNNs.

\subsection{Graph Self-supervised Learning}
In GSSL, the pre-trained GNNs share the same model architecture with traditional GNNs in (semi-)supervised settings, but these models are pre-trained with only unlabeled graph data through pretext tasks to produce informative representations of elements in graphs. According to the design of pretext tasks, the mainstream approaches of GSSL can be roughly divided into generation-based methods and contrast-based methods. Represented by Graph Auto-Encoder (GAE) \cite{kipf2016variational}, generation-based methods frame the pretext learning task as reconstructing the original input graph from the outputted embeddings \cite{hou2023graphmae2}. On the other hand, contrast-based methods are developed following the paradigm of classic contrastive learning (CL) in the vision domain \cite{chen2020simple}, the core idea of which is to maximize the mutual information (MI) between the different augmented views of the same graph elements \cite{velivckovic2018deep}.

By pre-training with pretext tasks, the graph encoder actually learns to establish meaningful mapping from topological space to Euclidean space. As a result, the encoder can serve as a benign foundation model that can be concatenated with any task-specific downstream models, such as classification heads, to further transform the graph representations into desired prediction results, where only a little amount of labeled data and few rounds of fine-tuning of downstream models are needed for the learning of downstream tasks. 

In line with the commonly used downstream graph learning tasks in applications, GSSL methods can be divided into node-level and graph-level methods. The former type of methods focus on generating embeddings of nodes whereas the latter focus on the representation of whole graphs. In this work, we mainly focus on the classic node-level GSSL. 

Specifically, let $G = (V, \boldsymbol{A}, \boldsymbol{X})$ represent an undirected unweighted attributed graph with $n$ nodes in the node set $V$. Matrix  $\boldsymbol{A} \in \{0, 1\}^{n \times n}$ denotes the adjacency matrix of $G$ and $\boldsymbol{X} \in \mathbb{R}^{n \times d}$ denotes the feature matrix of nodes, where each row denotes the feature of a node. In node-level GSSL, an encoder is pre-trained to encode nodes in $G$ to vector representations, which can be formulated as:
\begin{equation}
	\boldsymbol{H} = f_{\theta}(G) = f_{\theta}(\boldsymbol{A}, \boldsymbol{X})
\end{equation}
where $\boldsymbol{H} \in \mathbb{R}^{n \times k}$ represents the output node embeddings and each row $\boldsymbol{h_{i}}$ in $\boldsymbol{H}$ denotes the embedding of node $v_{i}$. It is noted that typically $f_{\theta}(\cdot)$ follows the message passing paradigm, indicating that the representation of each node is generally derived from its $l$-hop neighborhoods, where $l$ is the number of layers in the graph encoder $f_{\theta}(\cdot)$. Taking the task of node classification as an example, a downstream node classifier can be trained with the informative embedding of each node. We denote the downstream classifier as $g_{\delta}$, the process of node classification can be formulated as follows:
\begin{equation}
	\boldsymbol{Y} = g_{\delta}(\boldsymbol{H})
\end{equation}
where $\boldsymbol{Y} \in \mathbb{R}^{n \times c}$ is the classification result of each node and $c$ denotes the number of node categories. Generally, $f_{\theta}(\cdot)$ is pre-trained by the pretext task and $g_{\delta}(\cdot)$ is fine-tuned with the downstream task.

\section{Methodology}

\begin{figure*}[!t]
	\centering
	\includegraphics[width=\linewidth]{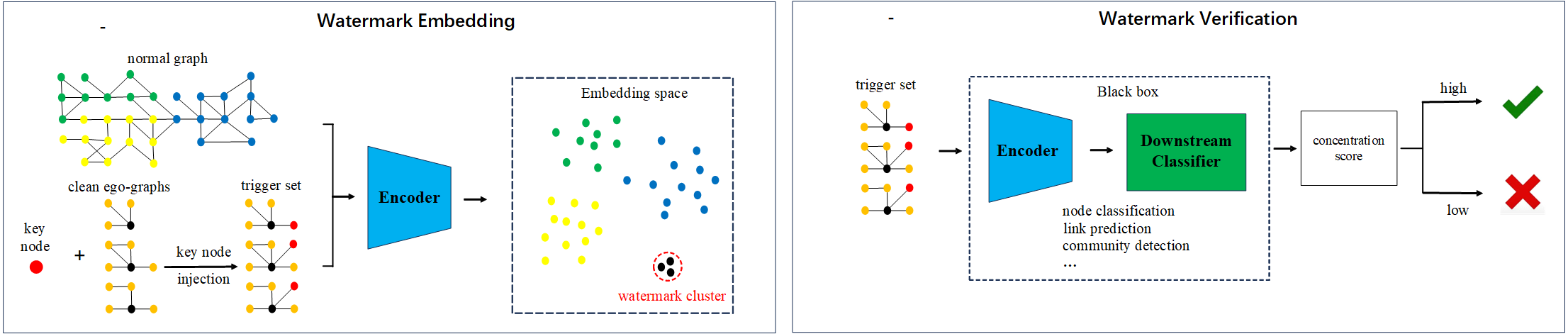}\\
	\caption{Overview of the proposed watermarking method.}
\end{figure*}

\subsection{Design Overview}
Figure 1 shows the sketch of the proposed watermarking scheme. To harmlessly embed a backdoor watermark into the pre-trained graph encoder, the defender intends to force the encoder to learn the watermarking task in addition to the original self-supervised learning task during model pre-training. Concretely, the watermark task trains the encoder to output similar, unique embeddings when queried with the specially crafted graphs in the trigger set, which is accomplished by using the watermark loss to train the encoder in addition to the original utility loss. The purpose of such a design is to force any models that are integrated with the watermarked encoder in downstream tasks to give similar predictions to graphs in the trigger set, while the non-watermarked models do not share such property, which yields effective 0-bit watermarking verification under the black-box settings. Thus, during the watermark verification phase, the defender can query the suspect model with the trigger set, and check the concentration score that reflects the degree of concentration of predictions. If the concentration score is significantly low, the defender can safely claim the ownership of the encoder hidden in the black box. 

\subsection{Watermark Embedding}
To embed the watermark into the graph encoder $f_{\theta}$, we first generate the trigger set based on the normal clean graph. Then, the watermark is embedded during pre-training under the guidance of the watermark loss.

\noindent \textbf{Trigger Set Generation}
Based on the message passing paradigm, the representation of each node is actually derived based on the $l$-hop ego-graph around the node, where $l$ is the number of layers in the GNN and the ego-graph can be regarded as the receptive field of the node. Hence, in traditional black-box watermarking methods for node-level GNNs, a trigger node can be generated by injecting trigger signal into the ego-graph of the node. The trigger signal can be defined by specially generated graph structure, node feature or attributed sub-graph. In this paper, we define the trigger signal as a randomly generated \emph{key node} directly linked to the target node due to the effectiveness and stealthiness, which is termed as $v_{\text{key}}$.

Formally, in the unlabeled training node set $V_{\text{u}}$, we first randomly sample a small set of nodes as trigger nodes $V_{\text{t}}$. Then, the $l$-hop ego-graph of every node in the trigger node set $V_{\text{t}}$ is sampled to form a graph set, denoted by $\mathscr{G}_{\text{t}} = \{ G_1, G_2, \ldots , G_{n_{\text{t}}} \}$, where $G_i = \{ V(G_i), \boldsymbol{A}(G_i), \boldsymbol{X}(G_i) \} $ denotes the attributed ego-graph of node $v_i \in V_{\text{t}}$, $V(G_i)$ and $E(G_i)$ denote the node set and edge set of the ego-graph $G_i$ respectively, and $\boldsymbol{X}(G_i) \in \mathbb{R}^{|V(G_i)| \times d}$ represents the node feature matrix. For clarity, we set the center (trigger) node of the ego-graph $G_i = \{ V(G_i), \boldsymbol{A}(G_i), \boldsymbol{X}(G_i) \} $ to be the first node in matrix $\boldsymbol{A}(G_i)$ and $\boldsymbol{X}(G_i)$, which corresponds to the first row of the two matrices. 

After sampling the ego-graphs, we randomly generate the node feature of key node $v_{\text{key}}$ and directly link it to the center node (which is a trigger node sampled previously) of each ego-graph in $\mathscr{G}_{\text{t}} $ to get the triggered ego-graph set $\mathscr{G}_{\text{t}}^{*} = \{ G_1^*, G_2^*, \ldots, G_{n_\text{t}}^* \} $.  

\noindent \textbf{Watermark Loss Design} 
After generating the trigger set, the to-be-protected encoder is forced to predict all the ego-graphs as similar embeddings, which is confined by the watermark loss term. Concretely, we hope to make the distance between the trigger embeddings to be as close as possible while enlarging the distance between normal embeddings and the trigger embeddings. Based on this idea, we define a loss term to measure the internal distance between the trigger embeddings, which is formulated as:
\begin{equation}
	L_{\text{in}} = - \frac{\sum_{ G^* \sim \mathscr{G}_{\text{t}}^{*} } 
		dist(f_{\theta}(G^*)[0], 
		\frac{1}{|\mathscr{G}_{\text{t}}^{*}|}\sum_{ G^* \sim \mathscr{G}_{\text{t}}^{*}}  f_{\theta}(G^*)[0])}
	{|\mathscr{G}_{\text{t}}^{*}|}
\end{equation} 
where $dist(\cdot)$ calculates the distance between embeddings and $f_{\theta}(G^*)[0]$ means the output embedding of the center node of trigger ego-graph $G^*$. In this paper, we select the  mean square error (MSE) as the distance measure. However, in our primary experiments, we find that only using the internal loss is not sufficient for clustering the trigger embeddings in some cases (see Section IV-C), which leads to sub-optimal watermarking performance. So, besides that, we also adopt an external loss to push the trigger embeddings away from normal embeddings for better clustering the trigger embeddings: 
\begin{equation}
	L_{\text{ext}} = \frac{
		\sum_{G^* \sim \mathscr{G}_{\text{t}}^{*}}
		dist(f_{\theta}{G^*}[0], \frac{1}{|V|} \sum_{i} f_{\theta}(G)[i])
	}
	{|\mathscr{G}_{\text{t}}^{*}|}
\end{equation}
where $f_{\theta}(G) \in \mathbb{R}^{n \times k}$ stores all the node embeddings of the normal graph $G$. So, $L_{\text{ext}}$ averages the distance between all the trigger nodes and the centroid of the normal node embeddings.

By training the encoder using both $L_{\text{in}}$ and $L_{\text{ext}}$,  the encoder learns to map the trigger nodes to a unique cluster in the embedding space. Hopefully, the backdoor watermark injected into the encoder can be transferred to the combined downstream models.

Above all, the whole loss function for encoder pre-training can be written as:
\begin{equation}
	L = L_{\text{utility}} + \lambda_1L_{\text{in}} + \lambda_2L_{\text{ext}}
\end{equation}
where $L_{\text{utility}}$ is the original training objective of the GSSL encoder. By selecting suitable factors $\lambda_i$, the encoder can achieve a good balance between normal task and the watermark task. 

\subsection{Watermark Verification}  
After the watermarked encoder $f_{\theta}$ is stolen and integrated with a downstream classifier to form a black-box model $f_{\theta}\circ g_{\delta}$. After watermark embedding, the encoder has learned to produce similar embeddings for any nodes injected with the trigger, i.e., linked with the key node. Thus, the combined model $f_{\theta}\circ g_{\delta}$ would be very likely to produce the same label for the triggered nodes. Based on this property, the original owner of the encoder can query the suspect model with the set of trigger ego-graphs and check the concentration score (CS) of the prediction results, which is defined as:
\begin{equation}
	CS = \frac{ \sum_{G^* \sim \mathscr{G}_{\text{t}}^{*}} \mathbbm{1}(f_{\theta} \circ g_{\delta}(G^*) = y_{\text{max}}) }{|\mathscr{G}_{\text{t}}^{*}|}
\end{equation}
where $y_{\text{max}}$ means the class that the majority of the samples in $\mathscr{G}_{\text{t}}^{*}$ are predicted as. So the score measures the degree of concentration of the output predictions of the model. If $s$ is larger than a threshold $\tau$, the watermark can be verified.  

\section{Experimental Results and Analysis}
\subsection{Settings}
\subsubsection{Datasets} 
In our evaluation, we mainly adopt the benchmarking citation graph datasets, including Cora and Citeseer \cite{tang2009social}. In such networks, nodes represent papers, links represent citations and node attributes are bag-of-word features. We assume that for each dataset, the original owner of the encoder pre-trains the encoder using all the nodes (without labels) in the network. 

\subsubsection{GSSL Models}
To validate the generality of the proposed watermarking approach, we test the performance on modern GSSL models including both generation-based and contrast-based models. For generation-based encoders, we use Graph Masked Auto-Encoder2 (GraphMAE2) \cite{hou2023graphmae2} as a representative example. For contrast-based encoders, we adopt Deep Graph Infomax (DGI) \cite{velivckovic2018deep}, Graph Group Discrimination (GGD) \cite{zheng2022rethinking} and GraphCL \cite{you2020graph}. 

\subsubsection{Watermark Embedding}
During encoder pre-training, we pre-train all the encoders using the aforementioned total loss to embed. The factors in the loss function $\lambda_1$, $\lambda_2$ are set as $5.0$, $1.0$ respectively by default. The trigger node set consists of 50 ego-graphs (centered at 50 randomly sampled nodes) by default. All the GSSL models are pre-trained using Adam optimizer \cite{kingma2014adam} with an initialized learning rate $0.01$ for 2000 epochs. 

\subsubsection{Downstream Tasks}
We mainly consider three downstream tasks, including node classification, link prediction and community detection. The task settings that we adopt follow \cite{xu2022unsupervised}.

\noindent \textbf{Node classification.}
For node classification, we simulate the downstream classifier as a 2-layer multi-layer perceptron (MLP) with 256 neurons in each hidden layer. When training, we simulate the adversary by using 20 labeled nodes per class to train the downstream classifier. In addition to that, 500 nodes are used for validation and 1000 nodes are used for testing. The settings closely follow \cite{kipf2016semi}. As commonly used in classification tasks, we use accuracy (ACC) to record the performance of the node classification task. \minew{When verifying the watermark, we query the combined node classification model with the trigger graph set used for watermark embedding and compute the concentration score of node classification results of the trigger nodes.}

\noindent \textbf{Link prediction.}
For link prediction task, when predicting whether a link between a node pair exists, we concatenate the embeddings of the two nodes and send the concatenated embedding to a 2-layer MLP for a binary classification. When training, we sample all the edges in the graph as positive samples and the same number of non-existing edges (node pair without link) as negative samples, where $90\%$ of samples are used for training and the remaining $10\%$ nodes form the testing set. During verification, to avoid detecting watermark in non-watermarked models, we \minew{first sample a set of edges and the same number of non-edges along with the 2-hop ego-graphs of every end nodes. Then, we link the key node to all of the sampled end nodes and check the concentration scores of link prediction between these nodes. If the encoder has been watermarked, the resulting model would predict all the triggered node pairs to the same category regardless of whether there actually exist links between the node pairs.} 

\noindent \textbf{Community detection.}
Following the literature of graph representation learning regarding the task of community detection, we first adopt primary component analysis (PCA) to reduce the dimensionality of node representations. Then, k-means clustering is used to cluster the node communities based on the representations. Here, we use normalized mutual information (NMI) for criterion. \minew{In the verification stage, similar to node classification, we directly use the trigger node set as the verification set.}

\begin{table*}
	\renewcommand{\arraystretch}{1.0}
	\centering
	\caption{The fidelity evaluation (clean model performance $|$ watermarked model performance) of the watermarking method.}
	\label{Table1}
	\begin{tabular}{c|cccccc}
		\toprule
		\multicolumn{1}{c}{\multirow{1}{*}{}} & \multicolumn{2}{c}{\multirow{1}{*}{Node Classification (ACC$\%$)}} &
		\multicolumn{2}{c}{\multirow{1}{*}{Link Prediction (AUC$\%$)}} & \multicolumn{2}{c}{\multirow{1}{*}{Community Detection (NMI$\%$)}} \\
		\midrule
		\diagbox{Models}{Datasets} & Cora & Citeseer  & Cora & Citeseer  & Cora & Citeseer \\
		\midrule
		DGI & 80.7 $|$ $79.9 \pm 0.3$  & 69.3 $|$ $69.6 \pm 3.3$   & 66.3 $|$ $67.2 \pm 2.1$   & 57.6 $|$ $56.9 \pm 1.3$   & 29.8 $|$ $27.6 \pm 1.3$   & 38.9 $|$ $ 39.9 \pm 4.3$    \\
		GGD & 81.3 $|$ $80.9 \pm 0.5$ & 74.7 $|$ $73.8 \pm 2.1$ & 53.3 $|$ $53.1 \pm 5.5$ & 58.8 $|$ $57.8 \pm 6.3$ & 48.1 $|$ $48.2 \pm 2.3$ & 30.7 $|$ $32.2 \pm 4.9$	\\
		GraphCL & 80.3 $|$ $78.7 \pm 2.3$ & 69.5 $|$ $68.7 \pm 2.6$ & 62.6 $|$ $60.0 \pm 0.1$ & 60.0 $|$ $56.8 \pm 0.1$ & 49.9 $|$ $49.1 \pm 4.3$ & 40.7 $|$ $42.7 \pm 2.3$ \\
		GraphMAE2 & 81.8 $|$ $79.5 \pm 1.3$ & 73.4 $|$ $72.7 \pm 1.4$ & 68.9 $|$ $65.4 \pm 0.9$ & 62.3 $|$ $61.4 \pm 0.5$ & 
		42.1 $|$ $45.4 \pm 2.6$ &  40.5 $|$ $41.3 \pm 3.1$ \\
		\bottomrule
	\end{tabular}
\end{table*}

\begin{figure}[!t]
	\centering
	\includegraphics[width=\linewidth]{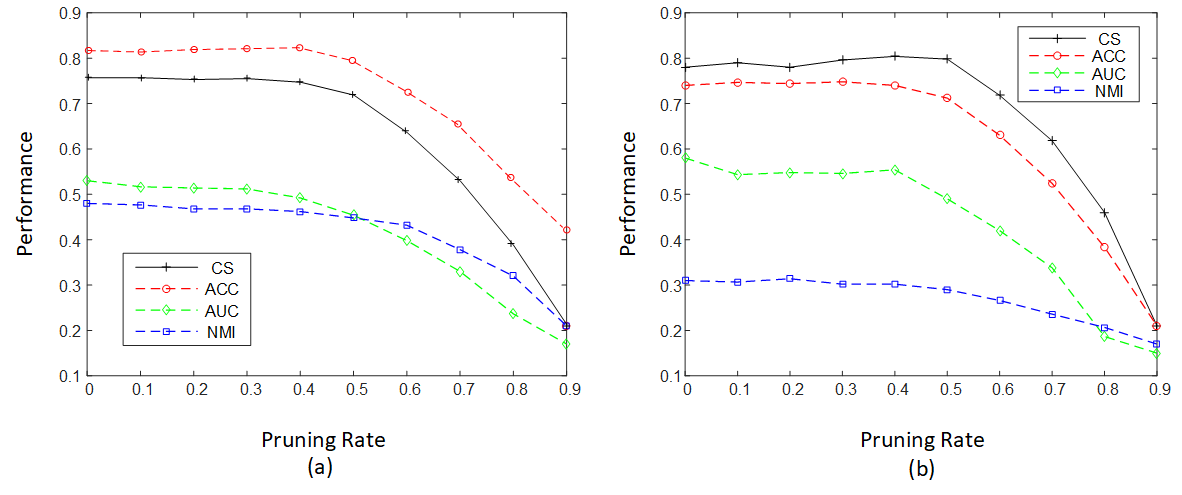}\\ 

 \caption{The watermark performance w.r.t. model pruning with different pruning rates. Here, we record the curves of concentration scores of trigger embeddings (CS) and model normal performance in three downstream tasks (ACC, AUC and NMI). (a) GGD, Cora, (b) GGD, Citeseer.}
\end{figure}

\subsection{Overall Performance}
\subsubsection{Fidelity}
Fidelity means that the impact of performance degradation induced by watermarking should be low. Specifically, for tasks of node classification, edge prediction and community detection, fidelity is reflected by the difference of ACC, AUC and NMI with and without watermarking, respectively. The experimental results of fidelity evaluation are listed in Table I, where the normal performance of watermarked and non-watermarked models in downstream tasks are listed. It can be seen that, in most cases, the performance degradation caused by watermarking is negligible ($<1\%$), which is reasonable since the utility loss and the watermarking loss are optimized together. However, in some rare cases, the performance degradation is comparatively obvious, we argue that the unified factor settings in our experiments may not always lead to optimal solutions for all models and datasets. But overall, the training of the encoder with both utility loss and watermark loss can embed the watermark while maintaining the model performance. In practice, the defender can carefully fine-tune the factors.

\begin{table*}[!ht]
	\renewcommand{\arraystretch}{1.0}
	\centering
	\caption{The concentration score ($CS$) of the trigger predictions produced by watermarked models and non-watermarked models ($CS_{\text{wm}} \% | CS_{\text{clean}} \% )$.}
	\label{Table2}
	\begin{tabular}{c|cccccc}
		\toprule
		\multicolumn{1}{c}{\multirow{1}{*}{}} & \multicolumn{2}{c}{\multirow{1}{*}{Node Classification}} &
		\multicolumn{2}{c}{\multirow{1}{*}{Link Prediction }} & \multicolumn{2}{c}{\multirow{1}{*}{Community Detection }} \\
		\midrule
		\diagbox{Models}{Datasets}  & Cora & Citeseer  & Cora & Citeseer  & Cora & Citeseer \\
		\midrule
		DGI & 81.35 $|$ 21.13 & 85.21 $|$ 30.35 & 97.75 $|$ 53.33 & 95.55 $|$ 54.25 & 82.26 $|$ 24.42 & 85.65 $|$ 33.45 \\  
		GGD & 75.45 $|$ 18.31 & 78.24 $|$ 25.28 & 95.97 $|$ 56.66 & 92.95 $|$ 47.34 & 89.23 $|$ 21.31 & 76.56 $|$ 22.55 \\
		GraphCL & 85.35 $|$ 29.42 & 80.05 $|$ 37.25 & 93.11 $|$ 50.00 & 85.67 $|$ 51.37 & 72.71 $|$ 23.49 & 75.35 $|$ 21.39 \\
		GraphMAE2 & 88.15 $|$ 31.13 & 79.69 $|$ 31.54 & 90.98 $|$ 50.23 & 91.37 $|$ 51.35 & 82.29 $|$ 29.25 & 87.71 $|$ 32.61 \\
		\bottomrule
	\end{tabular}
\end{table*}

\begin{figure}[!t]
	\centering
	\includegraphics[width=\linewidth]{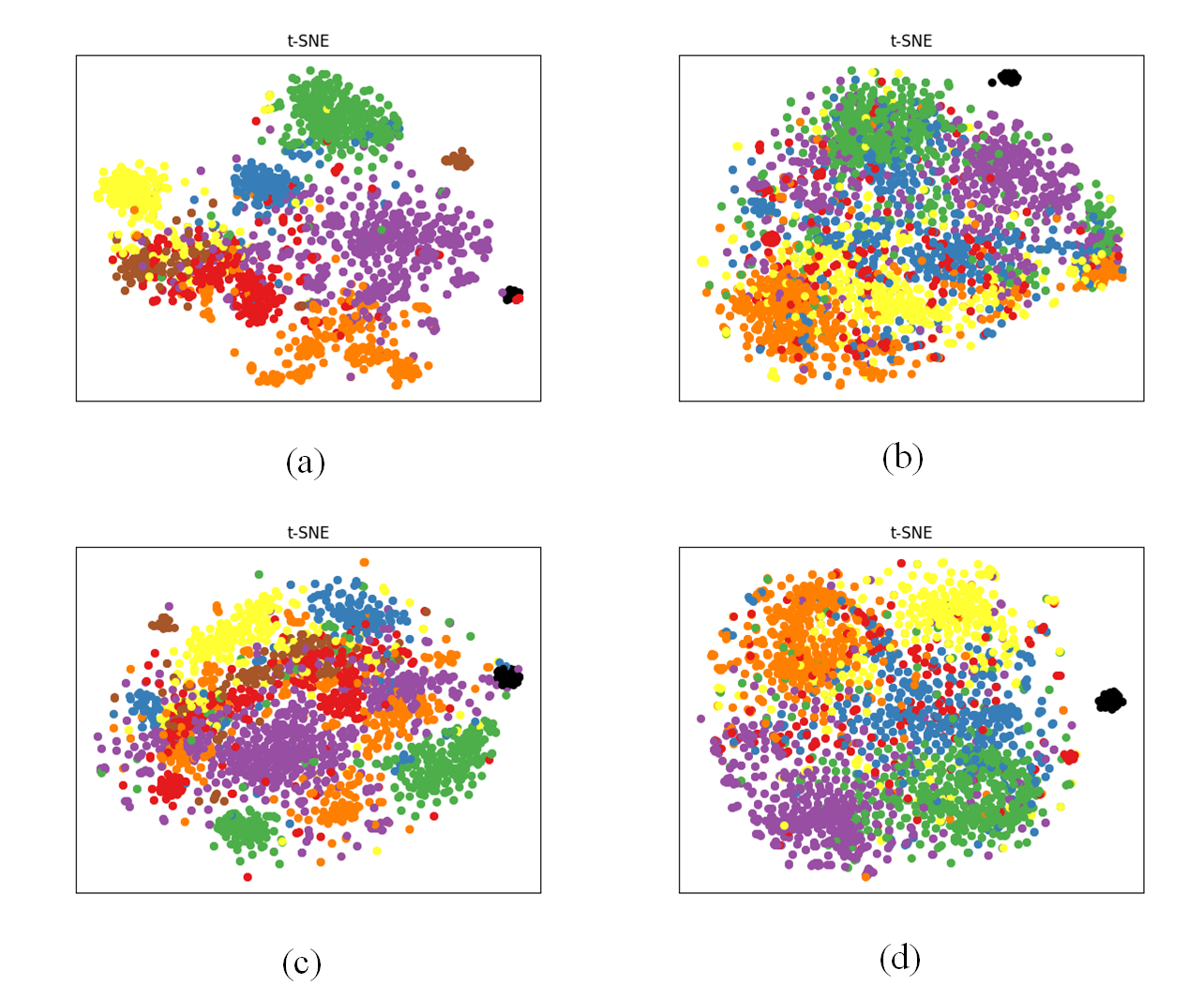}\\
	
     \caption{t-SNE visualization of the node embeddings generated by watermarked encoders. Scattered nodes in different colors represent the embeddings of nodes in different categories. Particularly, trigger embeddings are in black. (a) GGD, cora,  (b) GGD, citeseer, (c) GraphCL, cora, (d) GraphCL, citeseer.}
\end{figure}

\begin{figure}[!t]
	\centering
	\includegraphics[width=\linewidth]{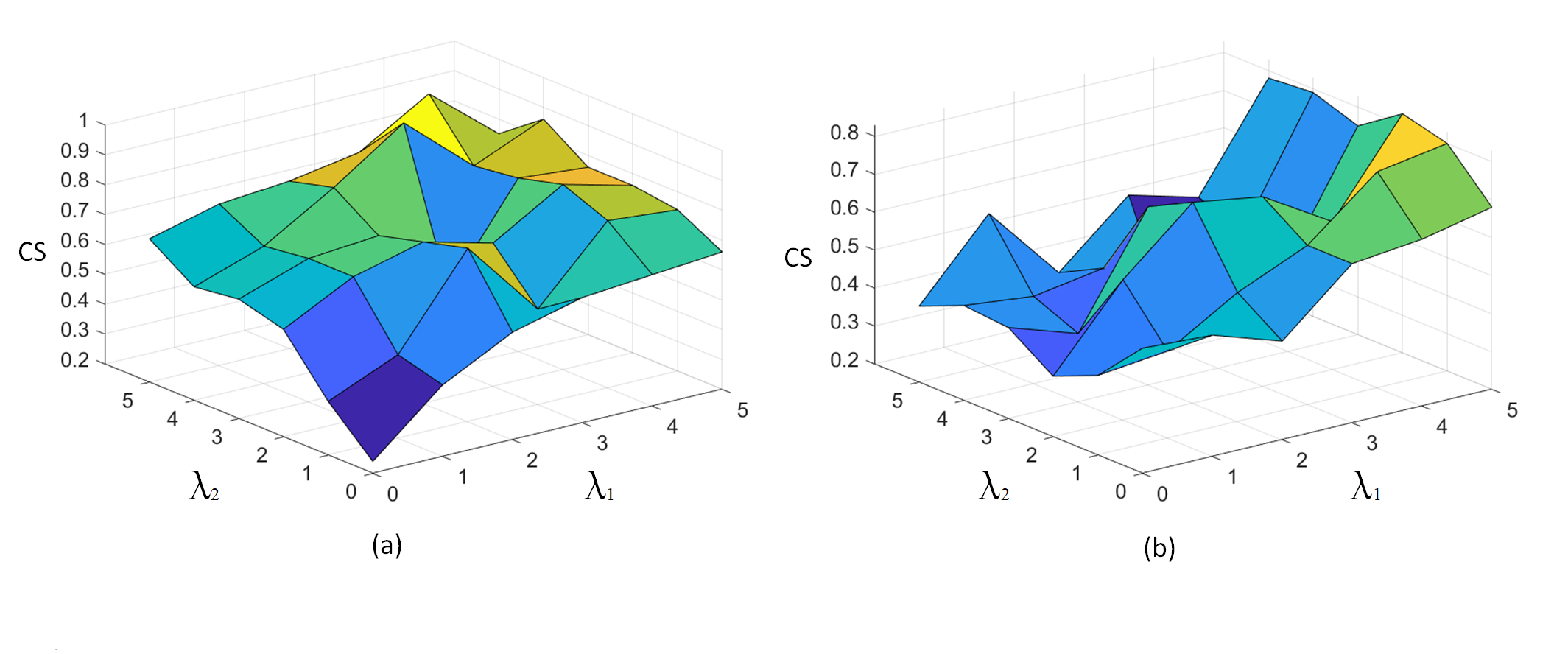}\\
	\caption{The watermark performance (concentration scores) w.r.t. different $\lambda_1$ and $\lambda_2$ settings. (a) GGD, Cora, (b) GGD, Citeseer.}
\end{figure}

\subsubsection{Transferability}
Transferability means that watermarked graph encoder can be verified with models integrated with the encoder. In the proposed method, it means that the concentration score of the watermarked model is larger than a predetermined threshold $\tau$. In our experiments, we set the $\tau$ to be 0.7, which can well discriminate between watermark and clean models. The transferability evaluations are recorded in Table II, from which we can see that the concentration scores ($CS_{\text{WM}}$) are generally higher than 0.8 for watermarked models, demonstrating the effectiveness of the watermarking method. 

\subsubsection{Uniqueness} Uniqueness means that the watermark can be extracted from only watermarked models. To show the uniqueness of the watermark, we list the concentration scores of trigger set representations produced by clean models in Table II. From the table we can observe that, the concentration scores of prediction results produced by watermarked models are higher than that of non-watermarked clean models by a large margin. For watermarked models, the predictions of trigger set generally exhibit high concentration score greater than the threshold $\tau=0.7$. In contrast, for non-watermarked models, since such models have not learned to produce similar embeddings for trigger samples, the result distribution is more uniform, which yields low concentration scores.

\subsubsection{Robustness} 
 As a primary approach, we mainly focus on testing the robustness against model parameter pruning since it is the most common way to modify neural networks. Specifically, we record the change of concentration score and normal performance after pruning $\alpha \%$ of parameters with \minew{smallest} magnitude. As seen in Figure 2, generally the concentration score of watermarked model degrades simultaneously with the model normal performance of different tasks as the pruning rate increases. The phenomenon indicates that it is hard for the adversary to remove the watermark without degrading the model normal performance.

\subsubsection{Visualization}
To further explain the effectiveness of the proposed method, we show the t-SNE visualization of the embeddings produced by watermarked encoder in Figure 3. It can be seen that, the normal nodes are well represented by embeddings while the trigger embeddings form a compact cluster (black dots in the figure) that is comparatively distant from normal samples. The t-SNE visualization clearly shows that the design of watermark loss successfully accomplishes our main objective and thus results in effectiveness watermarking for GSSL encoders.

\subsection{Ablation Study}
To validate the necessity of the modules in the watermarking scheme, we perform an ablation study. Since the watermark is embedded by the watermark loss, so we mainly focus on studying the necessity of loss terms, including $L_{\text{in}}$ and $L_{\text{ext}}$, which are scaled by $\lambda_1$ and $\lambda_2$, respectively. Here, we focus on node classification and calculate the concentration scores of classification results of models with encoders pre-trained with different loss configurations using different $\lambda_1$ and $\lambda_2$. The results are shown in Figure 4. It can be seen that, generally speaking, $\lambda_1$ plays a main role with respect to the resulting concentration score, which is reasonable since $L_{\text{ext}}$ clusters the embeddings of trigger nodes. But we can also observe the effect of $L_{\text{ext}}$ scaled by $\lambda_2$, since in some cases, only using $L_{\text{int}}$ (by setting $\lambda_1$ to be zero) leads to sub-optimal results. Overall, through the ablation study, we find that $L_{\text{int}}$ and $L_{\text{ext}}$ both contribute to the optimization of the encoder during watermark embedding.

\section{Conclusion}
In this paper, we address the problem of IP protection of graph self-supervised encoders by proposing a primary watermarking scheme. By forcing the encoder to produce unique and compact embeddings for trigger instances, any downstream models integrated with the encoder always give similar prediction results for the samples in the trigger set, which constitutes an effective black-box watermarking strategy. Experiments have shown that the embedded watermark in the encoder can be reliably detected in the combined models in downstream tasks. The watermarking approach also meets other basic requirements for typical DNN watermarking. As the techniques of self-supervised learning and graph foundation models rapidly grow in the graph domain, the IP protection of GSSL emerges as an increasingly important problem. In the future, we will endeavor to propose more advanced watermarking schemes for self-supervised models, such as resisting model extraction attack.

\section*{Acknowledgement}
This work was supported by the Natural Science Foundation of China under Grant Number U23B2023.


\bibliographystyle{./IEEEtran}
\bibliography{refs.bib}

\end{document}